\documentclass{article}
\usepackage{ijcai25}

\usepackage{times}
\usepackage{soul}
\usepackage{url}
\usepackage[hidelinks]{hyperref}
\usepackage[utf8]{inputenc}
\usepackage[small]{caption}
\usepackage{graphicx}
\usepackage{amsmath}
\usepackage{amsthm}
\usepackage{booktabs}
\usepackage{algorithm}
\usepackage{algorithmic}
\usepackage{amssymb}
\usepackage[dvipsnames]{xcolor}

\usepackage{multirow}
\usepackage{array}
\usepackage{longtable}
\usepackage{subfigure}

\title{GOT4Rec: Graph of Thoughts for Sequential Recommendation}

\author{
Zewen Long$^{1,2}$
\and
Liang Wang$^{1,2}$\and
Shu Wu$^{1,2}$\and
Qiang Liu$^{1,2}$\and
Liang Wang$^{1,2}$\\
\affiliations
$^1$NLPR, MAIS, Institute of Automation, Chinese Academy of Sciences\\
$^2$School of Artificial Intelligence, University of Chinese Academy of Sciences
\emails
\{zewen.long, liang.wang\}@cripac.ia.ac.cn,
\{shu.wu, qiang.liu, wangliang\}@nlpr.ia.ac.cn
}

\begin{document}

\maketitle

\begin{abstract}
With their vast open-world knowledge and reasoning abilities, large language models (LLMs) have become a promising tool for sequential recommendation.
Researchers have explored various methods to harness these capabilities, but most existing approaches rely on simple input-output prompting, failing to effectively bridge the gap between LLMs' general knowledge and the specific needs of recommendation tasks. While reasoning strategies like chain-of-thought (CoT) have been introduced to enhance performance, they often produce inaccurate recommendations due to underutilized user preference information and insufficient reasoning depth.
To address these challenges, we propose GOT4Rec, a novel sequential recommendation method leveraging the graph of thoughts (GoT) reasoning strategy. Our method focuses on three key types of information in user histories: short-term interests, long-term interests and collaborative information from other users. It enables LLMs to reason independently and generate recommendations, subsequently aggregating results to derive final items. This method allows LLMs, with enhanced reasoning capabilities, to better utilize the user sequence information, producing more accurate recommendations and comprehensive explanations.
Extensive experiments on real-world datasets demonstrate the effectiveness of GOT4Rec, outperforming existing state-of-the-art baselines with an average improvement of 37.11\%. Our code is available at https://anonymous.4open.science/r/GOT4Rec.
\end{abstract}

\section{Introduction}

Sequential recommendation has long been a significant research field, with numerous methods proposed to explore chronological dependencies within user sequences \cite{kang2018self,s3rec}. Despite considerable advancements, they remain constrained by the limited knowledge available from datasets. To overcome this, it is crucial to integrate real-world knowledge into sequential recommendation models, enabling them to more effectively comprehend and reason about preference patterns within user behavior sequences \cite{HouMZLDW22,LiWLFSSM23}.

\begin{figure}[t]
    \centering
    \includegraphics[width=\linewidth]{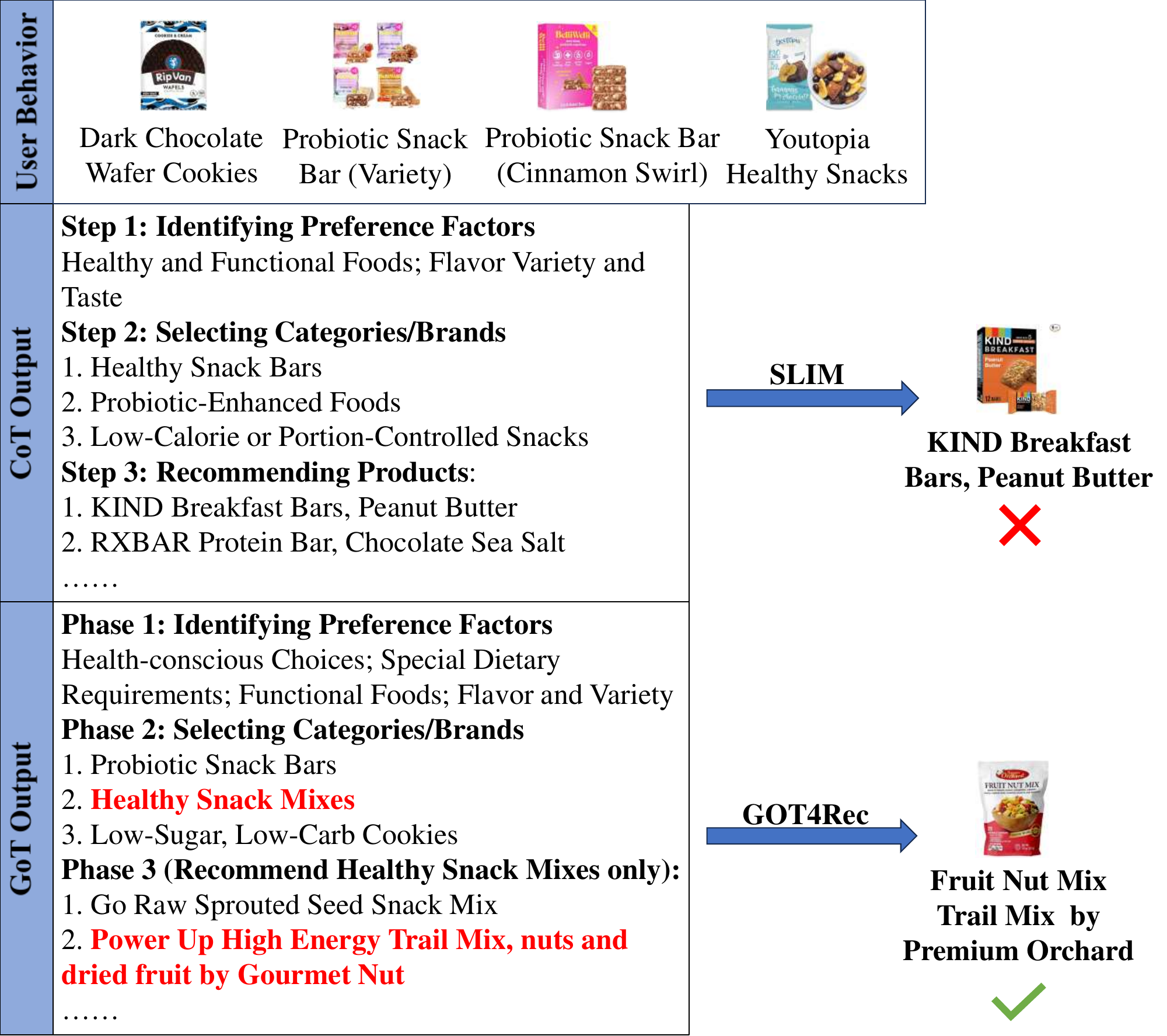}
    \caption{Comparison of the LLM output and predictions for the next item generated by SLIM and our proposed GOT4Rec.}
    \label{figure1}
\end{figure}

Large language models (LLMs) have recently attracted significant attention for their expansive open-world knowledge and advanced reasoning abilities. Consequently, numerous LLM-based sequential recommendation methods \cite{enhancer1,ranker4,ranker5} have emerged, aiming to capture user interests from interaction sequences while leveraging LLMs’ comprehensive real-world knowledge for recommendations. Many of these approaches rely solely on the input-output prompting paradigm, which underutilizes the reasoning potential of LLMs. As a result, it often produces task-irrelevant or inaccurate outcomes.
To address this problem, several studies have explored advanced reasoning strategies to boost LLMs’ performance in sequential recommendation. One example is SLIM \cite{wang2024can}, a knowledge distillation module which employs chain-of-thought (CoT) \cite{wei2022chain} prompting to enable step-by-step reasoning in sequential recommendation, transferring knowledge from a teacher model to a student model. However, it only utilizes the basic reasoning abilities of LLMs. This is insufficient for effectively processing the rich preference information embedded in user sequences, such as long-term and short-term interests or spatio-temporal patterns. Consequently, it often produces inaccurate recommendations. As illustrated in Figure \ref{figure1}, SLIM’s CoT-based reasoning focuses narrowly on predicting snack bars, while the true preference is a fruit nut mix, highlighting its limitations in handling diverse preference structures.

Therefore, the challenge lies in integrating diverse user preferences into a cohesive reasoning framework, turning sequential recommendation into a multi-faceted task requiring advanced reasoning capabilities. Studies have shown that simple approaches like CoT are inadequate for tasks that demand decomposition into sub-tasks \cite{got,cotnotgood}. In contrast, the graph of thoughts (GoT) method offers a more effective solution by decomposing the problem into more manageable components. Unlike CoT, which employs a single chain of thought, GoT models the reasoning process through networked reasoning. This approach allows GoT to break down the complex user preference reasoning tasks into smaller, more tractable sub-tasks, solve them individually, and then integrate the results to form a comprehensive solution. As illustrated in Figure \ref{figure1}, GoT reasons over multiple product categories that the user might be interested in: Probiotic Snack Bars, Healthy Snack Mixes and Low-Sugar, Low-Carb Cookies, enabling the LLM to independently generate recommendations within each category and combine them to correctly predict the ground truth item: fruit nut mix. While SLIM's recommendation remains focused on snack bars, showcasing its narrower approach.

To address the challenges mentioned above, this paper proposes GOT4Rec, a novel method that introduces the graph of thoughts (GoT) framework into the field of sequential recommendation. GOT4Rec optimally harnesses the reasoning capabilities of LLMs while effectively integrating multiple sources of information from user sequences. Specifically, we employ the GoT reasoning strategy to extract three critical sources of information from user sequences: short-term interests, long-term interests, and collaborative interests from other users with similar preferences. Unlike previous methods that treat the sequence as a whole,our approach considers multiple aspects of information and better utilizes the reasoning abilities of LLMs, leading to more accurate and interpretable recommendations. Our key contributions can be summarized as follows:

\begin{itemize}
    \item To the best of our knowledge, this work is among the first to explore the application of the graph of thoughts framework in the context of sequential recommendation.
    \item We introduce GOT4Rec, which optimally leverages the reasoning capabilities of LLMs to comprehensively capture and utilize the information within user sequences.
    \item Extensive experiments conducted on three datasets demonstrate that our method outperforms  neural and LLM-based sequential models as well as other LLM reasoning strategies.
    
\end{itemize}

\section{Related Work}

\subsection{Sequential Recommenders}

Traditional neural sequential recommendation systems aim to capture sequential dependencies in user behavior sequences to model dynamic user preferences, as seen in methods like GRU4Rec \cite{GRU4rec}. Techniques like self-attention and graph neural networks have further advanced the field \cite{kang2018self,s3rec,wu2019session,graph1,graph2}. These methods however, predominantly rely on sequence modeling capabilities and often inadequately incorporate textual information. Recently, research on transferable item representations \cite{HouMZLDW22,LiWLFSSM23} has gained attention. These approaches remain constrained by limited datasets and fail to fully leverage real-world knowledge. With the advent of LLMs, new potential are offered by leveraging LLMs' language understanding for recommendations. They can enhance features \cite{enhancer1,enhancer2,enhancer3} or act as rankers \cite{ranker1,ranker2,ranker3,ranker4,ranker5}. For example, ReLLa \cite{enhancer2} employs semantic user behavior retrieval to improve data quality and introduces retrieval-enhanced instruction tuning to enhance few-shot recommendation. LLMRank \cite{ranker5} formalizes the recommendation problem as a conditional ranking task and utilizes LLMs as zero-short rankers. Although these approaches are promising, existing research has not yet fully capitalized on the reasoning capabilities of LLMs.

\subsection{LLM Reasoning Strategies}

Numerous approaches have been proposed to exploit the reasoning capabilities of LLMs. Chain-of-thought (CoT) \cite{wei2022chain} introduces intermediate reasoning steps to enhance LLM performance, while Chain of Thought with Self-Consistency (CoT-SC) \cite{cot-sc} refines this by generating multiple CoTs and selecting the best output. Tree of thoughts (ToT) \cite{yao2024tree} and graph of thoughts (GoT) \cite{got} further extend these methods by modeling reasoning as a tree or graph, respectively, to better generate and aggregate different thoughts, thereby improving complex  tasks. In the recommendation domain, SLIM \cite{wang2024can} utilizes a CoT-based knowledge distillation module to transfer the step-by-step reasoning capabilities from a teacher model to a student model. However, its reliance on basic CoT  limits its ability to fully capture the abundant sequential dependencies and intricate patterns within user interactions.

\begin{figure*}[!ht]
    \centering
    \includegraphics[width=\linewidth]{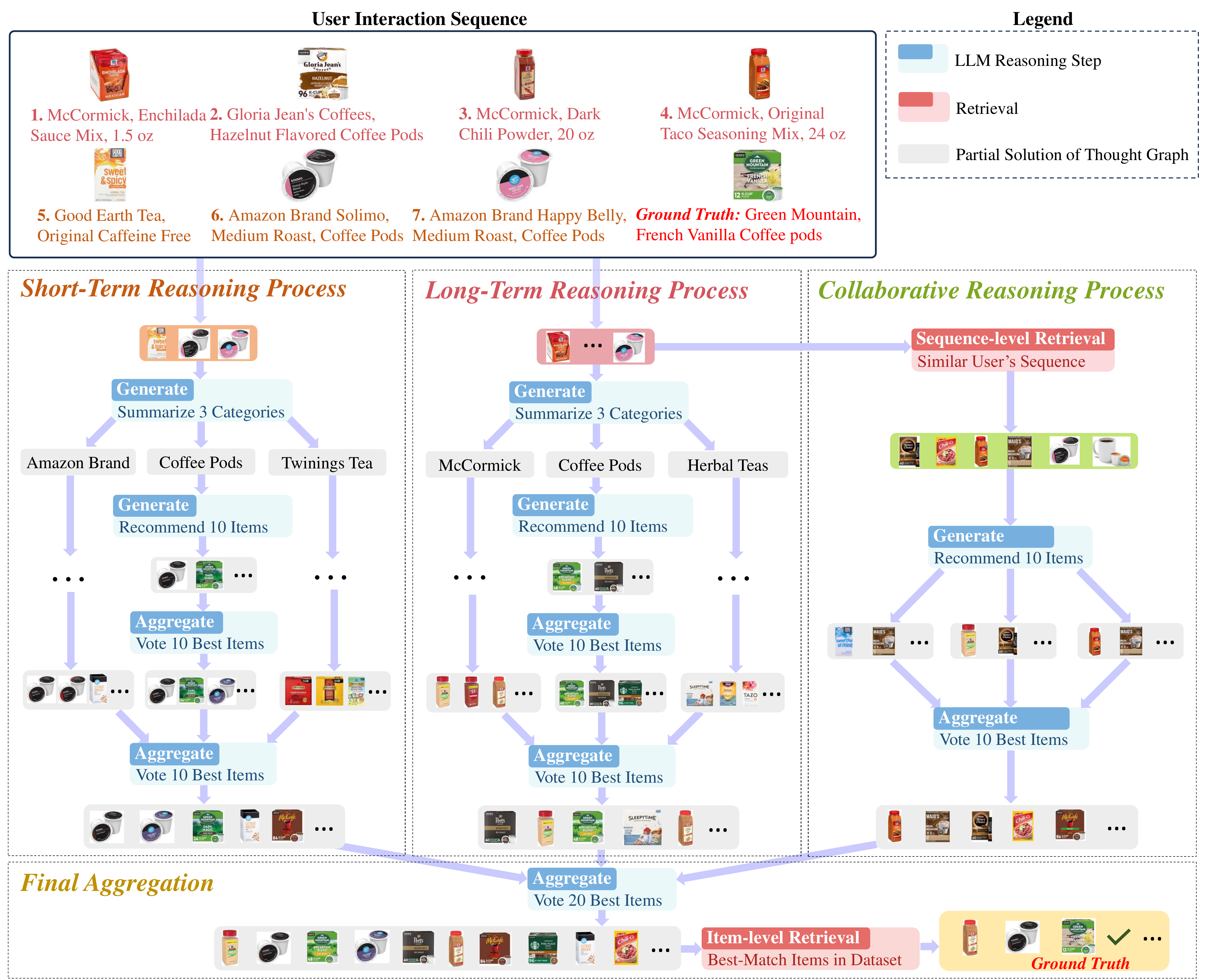}
    \caption{An example of graph decomposition in the proposed GOT4Rec method. The user interaction sequence is divided to facilitate different types of reasoning: the last few items are used for short-term preference reasoning, while the entire sequence informs long-term preference reasoning. Additionally, sequences from similar users are retrieved for collaborative reasoning. In the thought graph, these thoughts are generated and subsequently aggregated to capture and integrate the various aspects of information in the user interaction sequence.}
    \label{got4rec}
\end{figure*}

\section{The Proposed GOT4Rec Method}
In this section, we introduce GOT4Rec, a sequential recommendation method that fully leverages the reasoning capabilities of LLMs to capture the diverse information within user sequences. In the context of sequential recommendation, given a user $u$'s historical interaction sequence $S_{u} = \{i_{1}, i_{2}, ... , i_{n-1}\}$, the task is to predict the next item $i_{n}$ that the user is most likely to interact with.

\subsection{Overview of Thought Graph}
When interacting with LLMs, we input messages (prompts) and LLMs respond with generated outputs (thoughts). Building on the GoT framework \cite{got}, we model our GOT4Rec method as a tuple $(G, T)$, where $G$ represents the reasoning process and $T$ denotes the thought transformations. Specifically, we model the reasoning process as a directed graph $G = (V, E)$, where $V$ is the set of vertices and $E\subseteq V\times V$ is the set of edges. Each vertex represents a thought, encapsulating a solution to the current recommendation step along with other relevant global information. A directed edge $(t_{1}, t_{2})$ signifies a dependency between thoughts, indicating that thought $t_{2}$ is generated by LLMs based on $t_{1}$. 

In our method, we employ the generation transformation $T_{G}$ and aggregation transformation $T_{A}$ from GoT framework. The generation transformation generates one or more new thoughts based on an existing thought $v$. In this operation, new vertices and edges are generated: $V_{+} = \{v_{1+}, ..., v_{k+}\}$ and $E_{+} = \{(v, v_{1+}), ..., (v, v_{k+})\}$, where $v_{1+}, ..., v_{k+}$ are new thoughts generated by $v$. Reasoning steps such as CoT can be incorporated into this process. While the aggregation transformation combines multiple thoughts into a consolidated thought, reinforcing their strengths while mitigating their weaknesses. A new vertex $v_{+}$ is created in this operation: $V_{+} = \{v_{+}\}$ and $E_{+} = \{(v_{1}, v_{+}), ..., (v_{k}, v_{+})\}$, where $v_{1}, ..., v_{k}$ are the aggregated thoughts.

Figure \ref{got4rec} provides an example of graph decomposition in our GOT4Rec method. In the recommendation pipeline, the current user's interaction sequence $S_{u}$ is utilized as input. We then generate three key aspects of user preference information: short-term preference, long-term preference and collaborative preference, with detailed definitions provided in the subsequent subsection. With these preference information, the LLMs are enabled to reason and generate a list of top-$N$ items that best align with the user's current interests. Finally, these items are aggregated, and the LLMs vote to select the most probable items for recommendation.

\begin{figure}[t]
    \centering
    \includegraphics[width=\linewidth]{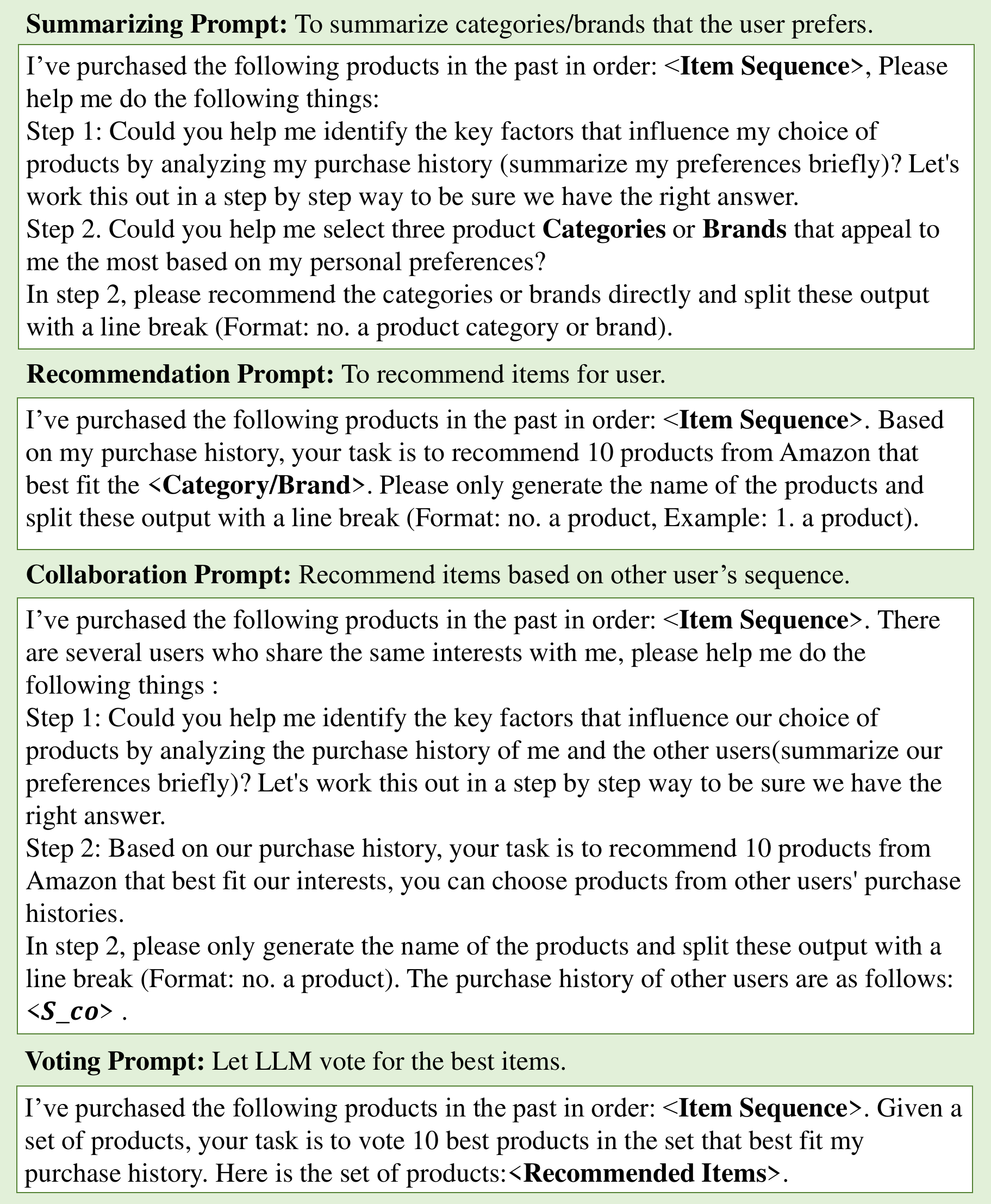}
    \caption{Prompt templates, including summarizing, recommendation, collaboration and voting prompts.}
    \label{prompt}
\end{figure}

\subsection{Short-Term \& Long-Term Reasoning Process}

Numerous studies have emphasized the importance of understanding both a user's dynamic short-term and stable long-term preferences in recommendation systems \cite{short-long2,short-long3}. In our GOT4Rec method, we enable LLMs to extract these two types of preferences separately and then synergize the most relevant aspects of each. To reason about short-term preferences, we select the last few interactions from the user sequence $S_{u}$ as $S_\text{short}$. We then apply and enhance the zero-shot CoT strategy from \cite{wang2024can} within the generation transformation $T_{G}$ to effectively capture the user's short-term preferences and identify three categories that the user is likely to favor. As illustrated in Figure \ref{prompt} as summarizing prompt, the generation transformation $T_{G}$ involves two key steps:

\begin{itemize}
    \item Step 1. Summarize user's short-term preferences based on the given $S_\text{short}$.
    \item Step 2. Based on the preferences in step 1, summarize three categories of products the user is inclined to prefer.
\end{itemize}

For each identified category, we prompt the LLMs to generate $N$ items the user is most likely interested in, repeating this process three times. The prompt template is provided in Figure \ref{prompt} as recommendation prompt. After generating the item sets, we utilize the aggregation transformation $T_{A}$, where the three item sets undergo a voting process by the LLMs. This results in a new set of $N$ items that are most likely to interest the user within each category. After aggergation, we have three final item sets, which are subjected to a final round of voting by the LLMs to determine the top-$N$ items $I_\text{short}$ that best represent the user's short-term preferences. The voting prompt is also provided in Figure \ref{prompt}.


To capture long-term preferences, we extend the input to include the entire user interaction sequence. Similar to the short-term reasoning process, we employ the zero-short CoT strategy within the generation transformation $T_{G}$, as shown in Figure \ref{prompt}. This allows LLMs to effectively differentiate and identify short-term and long-term preferences. After generating the relevant items, the final set of top-$N$ items $I_\text{long}$, representing the user's long-term preferences, is determined through the aggregation transformation $T_{A}$.

\subsection{Collaborative Reasoning Process}

Collaborative filtering is widely used in various recommendation scenarios, modeling users’ preference based on their past interactions. In our method, we leverage the all-mpnet-base-v2 \cite{all-mpnet-base-v2}, a sentence-transformers model that maps sentences to a vector space, to generate an embedding vector for the current user's interaction sequence, allowing us to retrieve a set of sequences $S_{co}$ from other users most similar to the current user's sequence. Details of this retrieval process are provided in the following section.

Once we have obtained the similar user sequences $S_\text{co}$, we employ the zero-short CoT strategy to generate three sets of top-$N$ items that the current user may be interested in. The two-step generation prompts, illustrated in Figure \ref{prompt} as collaboration prompt, are as follows:

\begin{itemize}
    \item Step 1. Summarize the shared preferences between the current user and other users based on the given $S_\text{co}$.
    \item Step 2. Based on the summarized preferences from Step 1, recommend $N$ items selected from $S_\text{co}$ that the current user is likely to prefer.
\end{itemize}

The three sets of items generated through this process are then aggregated and selected by the LLMs using the aggregation transformation $T_{A}$. This results in the final set of top-$N$ items $I_\text{co}$ that best reflect the collaborative preferences of users with similar interests to the current user.

At this point, we have obtained three distinct sets of items: $I_\text{short}$, which reflects the user's short-term preferences; $I_\text{long}$, which captures the user's long-term preferences; and $I_\text{co}$, which represents the collaborative preferences derived from other users. Finally, using the aggregation transformation $T_{A}$, we enable LLMs to synthesize these multiple sources of information to generate the final set of top-$N$ recommended items $I_\text{fin}$. The prompt template for this final aggregation is the same as the voting prompt.

\subsection{Multi-Level Retrieval Module}

To identify other users' sequences that share same interests with the current user and to assess how closely the recommended items match the ground truth in sampled datasets, we employ a two-level retrieval approach.

\paragraph{Sequence-level Retrieval.} As previously mentioned, when analyzing collaborative preferences, it is essential to retrieve sequences of other users who share similar interests with the current user. We choose to deploy the Mpnet \cite{all-mpnet-base-v2} model $f_\text{mpnet}$ due to its superior efficiency and discriminative power in encoding item titles compared to other encoding models such as BERT. Given a query sequence $S_{u} = \{i_{1}, ..., i_{m}\}$, the retrieved sequences set ${\mathcal{S}_{\mathcal{V}_\text{seq}}^{k}}$ can be obtained as follows:
\begin{equation}
    {\mathcal{S}_{\mathcal{V}_\text{seq}}^{k}}(i_t) \triangleq {\arg\max_{\boldsymbol{v} \in \mathcal{V}_\text{seq}}^k} sim(\frac{1}{m}\sum_{t=1}^{m}f_\text{mpnet}(i_{t}), \boldsymbol{v})
\end{equation}
where ${\mathcal{S}_{\mathcal{V}_\text{seq}}^{k}}$ is the set of top-$k$ sequences that share similar interests with the current user. $sim(\cdot,\cdot)$ denotes the computation of Euclidean distance, $V_\text{seq}$ represents the vector base containing the embeddings of all other sequences.

\paragraph{Item-level Retrieval.} The title of an item generated by LLMs may differ from the title of the ground truth in datasets, even though they refer to the same item. This discrepancy arises due to the limited and varied versions of items in the datasets (e.g., \textit{Witcher 3: Wild Hunt} versus \textit{Witcher 3: Wild Hunt Complete Edition}). To address this issue, we continue to utilize the $f_{mpnet}$ model to encode item titles into vectors. We then retrieve the most similar items from the vector base by computing the inner product of the query vector and all other vectors in the base. Specifically, for an item $i_\text{query}$ generated by LLMs, the retrieval process is as follows:
\begin{equation}
    {\mathcal{I}_{\mathcal{V}_\text{item}}^{k}}(i_\text{query}) \triangleq {\arg\max_{\boldsymbol{v} \in \mathcal{V}_\text{item}}^k} sim(f_\text{mpnet}(i_\text{query}), \boldsymbol{v})
\end{equation}
where ${\mathcal{I}_{\mathcal{V}_\text{item}}^{k}}$ represents the set of the retrieved top-$k$ items based on $sim(\cdot,\cdot)$, which denotes the computation of the inner product and $V_\text{item}$ is the vector base containing all item embeddings. In practice, we observed that in most cases, when the description of the query item is vague, the retrieved items are misaligned (e.g., when querying for \textit{The Witcher 3}, the top results include [\textit{Diablo III}, \textit{The Witch and the Hundred Knight - PlayStation 3}, \textit{The Witcher 3: Wild Hunt - Xbox One}]). To mitigate this, we retrieve the top-$K$ items within the ranked indices for each recommended item.

\subsection{Inference Latency and Volume Analyzation}

In Table \ref{tab:time}, we compare the inference lantency and volume between reasoning frameworks. Lantency is the number of hops to reach the final thought and volume refer to the number of preceding thoughts that have impacted the final thought \cite{got}. The time to generate a single thought is assumed to be $O(1)$. In CoT, reasoning follows a single sequential chain, resulting in both latency and volume scaling with $N$. CoT-SC employs $k$ independent chains, noting that these chains support parallel execution, both are reduced to $N/k$. For a $k$-ray ToT tree, both latency and volume are $\log_{k}N$. As for GoT, it is a $k$-ray tree combined with its inverted counterpart, resulting in a latency of $\log_{k}N$. For volume, since all intermediate thoughts contribute to the final thought by aggregating information from every step, it scales with $N$.

\begin{table}[htbp!]
\centering
\begin{tabular}{lll}
\hline
Framework & Latency & Volume \\ \hline
 Chain-of-Thought (CoT) & $N$ & $N$ \\
 Multiple CoTs (CoT-SC) & $N/k$ & $N/k$ \\
 Tree of Thoughts (ToT) & $\log_{k}N$ & $\log_{k}N$ \\ \hline
Graph of Thoughts (GoT) & $\log_{k}N$ & $N$ \\ \hline
\end{tabular}
\caption{Comparison of inference latency and volume for different reasoning frameworks.}
\label{tab:time}
\end{table}

Overall, by empowering LLMs to independently reason and generate thoughts and then aggregate the most relevant results, our method fully leverages the reasoning capabilities of LLMs in the sequential recommendation scenario. This approach enables LLMs to effectively capture and integrate different aspects of the extensive information within user sequences.

\section{Experiments}

\subsection{Experimental Settings}

\paragraph{Datasets.} We conduct experiments on three item categories from Amazon Reviews'23 dataset \cite{hou2024bridginglanguageitemsretrieval}: Video Games (\textbf{Games}), Grocery and Gourmet Food (\textbf{Food}) and Home and Kitchen (\textbf{Home}). Reviews are treated as user-item interactions, sequenced chronologically by timestamps. We focus on users with 6 to 20 interactions and filter out items with fewer than five interactions. Following \cite{wang2024can}, we randomly sample 3,000 users from each dataset three times. We employ the leave-one-out strategy for dataset division: the most recent interaction for testing, the second most recent interaction for validation and the remaining are used for training. Both training and validation segments are used as input during testing.

\begin{table*}[t]
\centering
\begin{tabular}{cc|ccc|cccc|cr}
\toprule
                 Dataset & Metric & SASRec & PALR & TALLRec & SLIM & CoT & CoT-SC & ToT  & GOT4Rec &  Improv.\\ \midrule
\multirow{6}{*}{Games} &HR@5 & \underline{0.0776} & 0.0630 & 0.0660 & 0.0602 &0.0644 & 0.0653 & 0.0489 &  \textbf{0.0894} &  15.21\% \\
                 &HR@10 & 0.0953 & 0.0893 & 0.0967 & 0.0977 &0.0988&  \underline{0.1013} & 0.0712  & \textbf{0.1167} &  15.20\%  \\ 
                  &HR@20 & 0.1227 & 0.1157 & 0.1233 & 0.1281& \underline{0.1347}  & 0.1253 & 0.1087  & \textbf{0.1361} &  1.04\%  \\
                  &NDCG@5  & \underline{0.0528} & 0.0385 & 0.0401 & 0.0380& 0.0408 & 0.0421 & 0.0304  & \textbf{0.0621} &  17.61\%  \\
                  &NDCG@10  & \underline{0.0586} & 0.0423 & 0.0497 & 0.0502&0.0519 &  0.0537 & 0.0377  & \textbf{0.0710} &  21.16\%  \\
                  &NDCG@20  & \underline{0.0655} & 0.0475 & 0.0562 & 0.0578 &0.0609&  0.0622 & 0.0471  & \textbf{0.0760} &  16.03\%  \\ \midrule
\multirow{6}{*}{Food} &HR@5  & 0.0335 & 0.0337 & 0.0390 & 0.0350&0.0396 &  \underline{0.0443} & 0.0273  & \textbf{0.0742} &  67.49\%  \\
                  &HR@10  & 0.0407 & 0.0487 & 0.0513 & 0.0517 &0.0581&  \underline{0.0597} & 0.0390  & \textbf{0.0972} &  62.81\%  \\
                  &HR@20  & 0.0549 & 0.0657 & 0.0747 & 0.0613 &0.0748&  \underline{0.0753} & 0.0533  & \textbf{0.1090} &  44.75\%  \\
                 &NDCG@5  & \underline{0.0286} & 0.0216 & 0.0243 & 0.0216& 0.0253 & 0.0276 & 0.0182  & \textbf{0.0492} & 72.03\%   \\
                  &NDCG@10  & 0.0309 & 0.0235 & 0.0284 & 0.0270 &0.0311&  \underline{0.0326} & 0.0219  & \textbf{0.0567} &  73.93\%  \\
                  &NDCG@20  & 0.0337 & 0.0267 & 0.0318 & 0.0295&0.0352 &  \underline{0.0366} & 0.0255  & \textbf{0.0597} &  63.11\%  \\ \midrule
\multirow{6}{*}{Home} &HR@5  & 0.0132 & 0.0130 & 0.0123 & 0.0123 &0.0118&  \underline{0.0133} & 0.0047  & \textbf{0.0192} &  44.36\%  \\
                  &HR@10  & 0.0179 & 0.0167 & 0.0183 & 0.0180&0.0177 &  \underline{0.0223} & 0.0100  & \textbf{0.0299} & 34.08\%   \\
                  &HR@20  & 0.0262 & 0.0213 & 0.0233 & 0.0213& 0.0230 &  \underline{0.0270} & 0.0147  & \textbf{0.0337} & 24.81\%  \\
                 &NDCG@5  & \underline{0.0097} & 0.0080 & 0.0081 & 0.0073& 0.0073 & 0.0080 & 0.0031  & \textbf{0.0122} &  25.77\%  \\
                  &NDCG@10  & 0.0105 & 0.0094 & 0.0096 & 0.0091 & 0.0093 &  \underline{0.0109} & 0.0049  & \textbf{0.0157} & 44.04\%   \\
                  &NDCG@20  & \underline{0.0134} & 0.0101 & 0.0113 & 0.0099&0.0106 &  0.0121 & 0.0060  & \textbf{0.0167} &  24.63\%  \\ 
\bottomrule
\end{tabular}
\caption{Recommendation performance. The best performance is highlighted in \textbf{bold} and the runner-up is highlighted by \underline{underlines}. Improvement indicates relative improvements over the best baseline in percentage.}
\label{results}
\end{table*}

\paragraph{Baselines.} To demonstrate the effectiveness of our proposed method, we select two groups of recommendation baselines. The first group comprises both neural and LLM-based sequential recommenders: 
\begin{itemize}
    \item \textbf{SASRec} \cite{kang2018self}: A self-attention based sequential model to capture user’s preferences.
    \item \textbf{PALR} \cite{palr}: A novel framework that integrates user behavior with LLMs to generate personalized recommendations.
    \item \textbf{TALLRec} \cite{tallrec}: An efficient fine-tuning framework that aligns LLMs with recommendation systems through a two-stage tuning process.
\end{itemize}

The second group contains models that utilize LLM reasoning strategies:
\begin{itemize}
    \item \textbf{SLIM} \cite{wang2024can}: A knowledge distillation module transferring the step-by-step reasoning capabilities in recommendation from a larger teacher model to a smaller student model.
    \item \textbf{Chain-of-Thought (CoT)} \cite{wei2022chain}: A reasoning approach for prompting which includes the intermediate steps of reasoning within the prompt.
    \item \textbf{Multiple CoTs (CoT-SC)} \cite{cot-sc}: A scheme in which multiple CoTs are generated, with the best one being selected as final result.
    \item \textbf{Tree of Thoughts (ToT)} \cite{yao2024tree}: A reasoning approach modeling LLM reasoning process as a tree.
\end{itemize}

\paragraph{Implementation Details.} We choose Llama3-8B-Instruct \cite{llama3} as the backbone model. SASRec and TALLRec are implemented based on their official code. We implemented PALR by ourselves and LLM reasoning strategies are integrated within the GoT framework. For SLIM, Llama3-8B-Instruct is used instead of ChatGPT to avoid high API costs. We retrieve the 10 most similar items for both GOT4Rec and baseline methods using the faiss library \cite{faiss}. Optimal hyper-parameters for all baselines are carefully selected to ensure the best performance.

\paragraph{Evaluation Metrics.} We evaluate performance with hit rate (HR) and normalized discounted cumulative gain (NDCG), reporting HR@K and NDCG@K for $K\in\{5, 10, 20\}$. Each recommended item is evaluated against all other items in the dataset. The average scores of three runs are reported.

\subsection{Overall Performance}
Table \ref{results} compares our GOT4Rec method with various neural sequential models and LLM reasoning strategies, leading to several key observations: 

\paragraph{Performance of Neural \& LLM-based Models.}
Neural and LLM-based sequential models perform relatively modest, though SASRec still outperforms LLM reasoning strategies in some cases, particularly due to its use of self-attention mechanisms which allow SASRec to effectively model users' historical behaviors by capturing the transition relationships between items. However, SASRec lacks the ability to comprehend semantic information, limiting its overall performance. As for PALR and TALLRec, their suboptimal performance stems from the absence of strong reasoning capabilities, which are critical for complex recommendation tasks. 
\paragraph{Performance of LLM Reasoning Methods.}
Among LLM reasoning methods, CoT-SC consistently achieves runner-up performance across most datasets, largely because it aggregates and selects the best result from multiple CoT paths, providing a more refined output. On the other hand, CoT, ToT, and SLIM show comparatively lower performance. SLIM, in particular, may suffer from reduced output diversity due to fine-tuning, while ToT's structural and reasoning path designs appear to be less suitable for sequential recommendation tasks.
\paragraph{Superiority of GOT4Rec.}
GOT4Rec achieves the state-of-art performances across all datasets. Notably, in the Food dataset, GOT4Rec achieves a relative improvement of 73.93\% over CoT-SC in terms of NDCG@10 and 67.49\% in terms of HR@5. These significant gains can be attributed to the nature of food product consumption, where users prefer consistent categories or brands and are strongly influenced by short-term needs. This aligns well with the strengths of GOT4Rec, which excels at capturing users' preferences for certain categories or brands and effectively integrating short-term preference information. This capability allows GOT4Rec to deliver highly relevant recommendations that closely match users' interests. These findings demonstrate the effectiveness of GOT4Rec in optimizing recommendation tasks by fully exploiting the advanced reasoning capabilities of LLMs and integrating various aspects of user information.

\newcommand{\blue}[1]{$_{\color{RoyalBlue}\downarrow #1}$}
\newcommand{\red}[1]{$_{\color{Salmon}\uparrow #1}$}

\begin{table*}[]
\centering
\begin{tabular}{cc|lllc}
\toprule
                Dataset  & Metric & $w/o$ Short-term & $w/o$ Long-term & $w/o$ Collaborative & GOT4Rec \\ \midrule
\multirow{6}{*}{Games} &HR@5& 0.0819\blue{8.39} & 0.0871\blue{2.57} & 0.0774\blue{13.42} & \textbf{0.0894}   \\
                    &HR@10& 0.1083\blue{7.20} & 0.1121\blue{3.94} & 0.1015\blue{13.02} & \textbf{0.1167}   \\
                  &HR@20& 0.1262\blue{7.27} & 0.1337\blue{1.76} & 0.1170\blue{14.03} & \textbf{0.1361}   \\
                  &NDCG@5& 0.0512\blue{17.55} & 0.0573\blue{7.73} & 0.0527\blue{15.14} & \textbf{0.0621}   \\
                  &NDCG@10& 0.0597\blue{15.92} & 0.0654\blue{7.89} & 0.0606\blue{14.65} & \textbf{0.0710}   \\
                  &NDCG@20& 0.0642\blue{13.53} & 0.0709\blue{6.71} & 0.0645\blue{15.13} & \textbf{0.0760}   \\ \midrule
\multirow{6}{*}{Food} &HR@5& 0.0591\blue{20.35} & \textbf{0.0867}\red{16.85} & 0.0687\blue{7.41} & 0.0742   \\
                &HR@10& 0.0867\blue{10.80} & 0.0933\blue{4.01} & 0.0880\blue{9.47} & \textbf{0.0972}   \\
                  &HR@20& 0.1003\blue{7.98} & 0.1063\blue{2.48} & 0.1013\blue{7.06} &  \textbf{0.1090}  \\
                  &NDCG@5& 0.0362\blue{26.42} & 0.0426\blue{13.41} & 0.0437\blue{11.18} & \textbf{0.0492}   \\
                  &NDCG@10& 0.0453\blue{20.11} & 0.0509\blue{10.23} & 0.0500\blue{11.82} & \textbf{0.0567}   \\
                  &NDCG@20& 0.0487\blue{18.43} & 0.0543\blue{9.05} & 0.0534\blue{10.55} &  \textbf{0.0597}  \\ \midrule
\multirow{6}{*}{Home} &HR@5& 0.0179\blue{6.77} & 0.0166\blue{13.54} & 0.0190\blue{1.04} & \textbf{0.0192}   \\
                &HR@10& 0.0284\blue{5.02} & 0.0292\blue{2.34} & 0.0270\blue{9.70} & \textbf{0.0299}   \\
                 &HR@20& 0.0322\blue{4.45} & 0.0318\blue{5.64} & 0.0309\blue{8.31} &  \textbf{0.0337}  \\
                 &NDCG@5& 0.0102\blue{16.39} & 0.0103\blue{15.57} & 0.0114\blue{6.56} & \textbf{0.0122}   \\
                  &NDCG@10& 0.0136\blue{13.38} & 0.0144\blue{8.28} & 0.0140\blue{10.83} &  \textbf{0.0157}  \\
                  &NDCG@20& 0.0146\blue{12.57} & 0.0150\blue{10.18} & 0.0149\blue{10.78} &  \textbf{0.0167}  \\ 
\bottomrule
\end{tabular}
\caption{Ablation analysis, conducted by retaining different components in GOT4Rec to form variants. The best performance is highlighted in \textbf{bold}. The performance difference (\%) with GOT4Rec is highlighted with {\color{RoyalBlue}{blue}} and {\color{Salmon}{orange}}. ``Short-term", ``Long-term", and ``Collaborative" denote short-term, long-term and collaborative reasoning steps, respectively.}
\label{tab:ablation}
\end{table*}

\subsection{Ablation Study}

We conducted an ablation study to analyze the impact of different components in the GOT4Rec model, with the results in Table \ref{tab:ablation}. Our GOT4Rec consistently outperforms the ablated variants, demonstrating that the full integration of users' preference information from the short-term, long-term, and collaborative components results in superior recommendation performance. The study also reveals that the importance of each component varies across different datasets, likely reflecting the unique characteristics of each dataset. 
For instance, collaborative information appears to be the most crucial component in Games dataset. This suggests that when purchasing gaming products, users prioritize categories or brands, making collaborative information from other users with similar interests particularly important.
In contrast, short-term preference plays a more significant role in the Food dataset, as users’ recent interactions are often influenced by immediate needs, cravings, or specific dietary goals. This aligns with our inference that users' recent preferences heavily influence their choices in food products. 
In the Home dataset, the impact of each component varies, but short-term preference has the least influence. This indicates that long-term preferences or collaborative insights are more critical when users make decisions about home-related items that typically involve more thoughtful consideration.

\subsection{Popularity Bias Analysis}

To assess recommendation novelty, we calculate EFD@10 and EPC@10 \cite{VargasC11}. EFD (Expected Free Discovery) represents the expected inverse collection frequency of relevant and seen recommended items. EPC (Expected Popularity Complement) measures the expected number of seen relevant recommended items that were previously unknown to the user. These two metrics provide a comprehensive assessment of diversity and novelty, helping to balance the recommendation of popular items with long-tail content. Table \ref{tab:bias} reports EFD@10 and EPC@10 metrics for CoT, SLIM and GOT4Rec, indicating that GOT4Rec outperforms the baselines in recommending long-tail items. These results support our claim that GOT4Rec mitigates popularity bias by capturing a wider range of information.

\begin{table}[htbp!]
\centering
\begin{tabular}{ccccc}
\hline
                Dataset  & Metric & CoT & SLIM & GOT4Rec \\ \hline
\multirow{2}{*}{Games} & EFD@10 & 4.8919 & 4.1292 & \textbf{5.0929} \\
                  & EPC@10 & 0.3885 & 0.3306 & \textbf{0.3998} \\ \hline
\multirow{2}{*}{Food} & EFD@10 & 6.4721 & 5.3517 & \textbf{6.8671} \\
                  & EPC@10 & 0.4750 & 0.3923 & \textbf{0.5035} \\ \hline
\multirow{2}{*}{Home} & EFD@10 & 4.1911 & 3.1141 & \textbf{4.6019} \\
                  & EPC@10 & 0.3110 & 0.2304 & \textbf{0.3450} \\ \hline
\end{tabular}
\caption{Popularity bias metrics, the higher score indicates higher recommendation novelty. The best score is highlighted in \textbf{bold}.}
\label{tab:bias}
\end{table}

\section{Conclusion}

In this paper, we propose GOT4Rec, a sequential recommendation method that optimally leverages the reasoning capabilities of LLMs to extract and integrate short-term, long-term, and collaborative user preferences utilizing the graph of thoughts (GoT) framework. Experiments on real-world datasets demonstrate that our GOT4Rec method outperforms existing neural sequential models and LLM reasoning strategies. Further analysis reveals that GOT4Rec effectively integrates multiple pieces of information contained within user sequences for superior recommendations.

\newpage


\bibliographystyle{named}
\bibliography{ijcai25}

\newpage
\appendix

\section{Dataset Statistics}

The detailed statistics of the three datasets used in our experiments are presented in Table \ref{datasets}, including the number of users, items, actions (the presence of a review is an action) and time spans covered by each dataset.

\begin{table}[!htbp]
\centering
\begin{tabular}{cccccc}
\hline
 Dataset & Users & Items & Actions & Time Span \\ \hline
 Games &  3,000 &  12259 &  26196 & 1999.11-2023.08 \\
 Food &  3,000 &  19256 &  29321 & 2003.11-2023.03 \\
 Home &  3,000 &  25457 &  29332 & 2001.01-2023.03 \\ \hline
\end{tabular}
    \caption{Datasets' average statistics (after preprocessing).}
    \label{datasets}
\end{table}

\section{LLM Generation Parameters}
In table \ref{tab:parameter}, we provide the detailed generation parameters of the LLMs used for generating responses to the given prompts. By outlining these parameters, we ensure reproducibility and offer insights into the configuration choices that impact model performance.

\begin{table}[!htbp]
\centering
\begin{tabular}{lc}
\hline
Parameter & Value \\ \hline
temperature & 0.6 \\
top-k & 40 \\
top-p & 0.8 \\
max sequence length & 4096 \\
presence penalty & 0.02 \\
frequency penalty & 0.02 \\
repetition penalty & 1.02 \\ \hline
\end{tabular}
\caption{LLM parameters for text generation.}
\label{tab:parameter}
\end{table}

\section{Popularity Bias Analysis}
\begin{figure}[t]
    \centering
    \subfigure[Popularity bias in Games dataset.]{
        \includegraphics[width=0.3\textwidth]{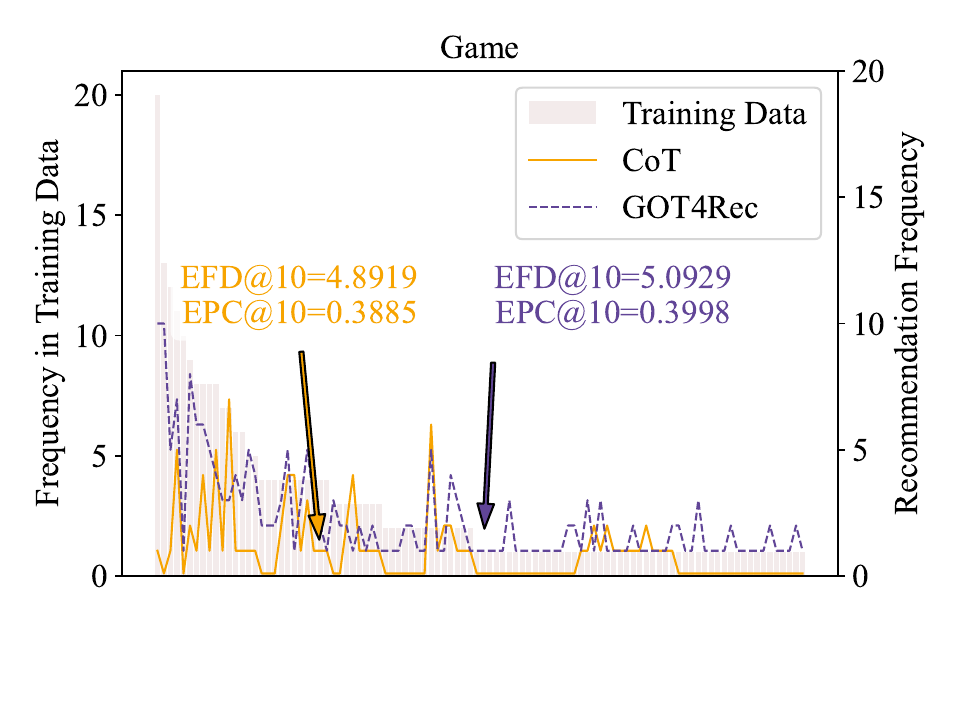}
        \label{fig:subfig1}
    }
    \subfigure[Popularity bias in Food dataset.]{
        \includegraphics[width=0.3\textwidth]{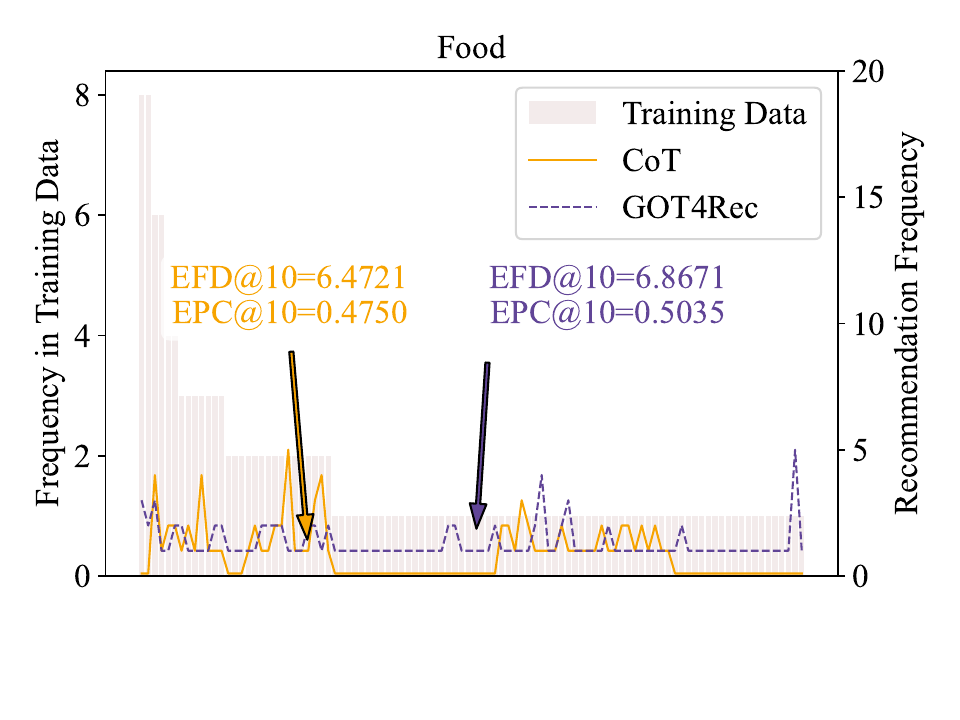}
        \label{fig:subfig2}
    }
    \subfigure[Popularity bias in Home dataset.]{
        \includegraphics[width=0.3\textwidth]{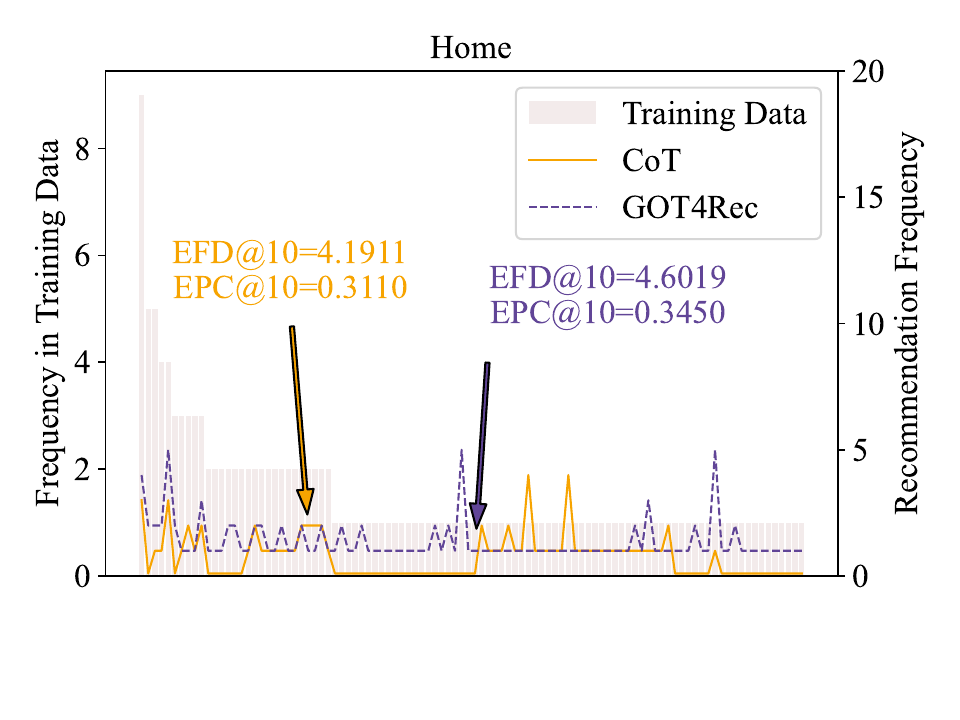}
        \label{fig:subfig3}
    }
    \caption{Analysis of popularity bias, items in the dataset are sorted by their frequency. Compared to CoT, GOT4Rec demonstrates a more consistent ability to recommend long-tail items.}
    \label{fig:bias}
\end{figure}

In Figure \ref{fig:bias}, we sort the items in all three datasets based on their frequency in the training set (i.e., popularity) and draw lines to illustrate each item's frequency in the results of CoT and GOT4Rec. This visualization highlights the distribution of recommendations across both popular and tail items. It is evident from the figure that GOT4Rec demonstrates a significant advantage in recommending tail items. Additionally, GOT4Rec achieves a broader coverage of items, indicating its ability to better capture and utilize the long-tail distribution.

\begin{table*}[t]
\centering
\begin{tabular}{cc|cllll}
\hline
                 Dataset & Metric & SOTA-Baselines & Llama3-8B & Llama3.2-3B & Gemma2-9B & Command-R-7B \\ \hline
\multirow{6}{*}{Games} & HR@5 &0.0776& \textbf{0.0894}\red{15.21} & 0.0623\blue{19.72} & 0.0655\blue{15.59} & 0.0604\blue{22.16} \\
                  & HR@10 &0.1013& \textbf{0.1167}\red{15.20} & 0.1041 \red{2.76}& 0.1152\red{13.72} & 0.1023\red{0.99} \\
                  & HR@20 &0.1347& 0.1361\red{1.04} & \textbf{0.1384}\red{2.75} & 0.1327\blue{1.48} & 0.1289\blue{4.31} \\
                  & NDCG@5 &0.0528& \textbf{0.0621}\red{17.61} & 0.0397\blue{24.81} & 0.0392\blue{25.76} & 0.0376\blue{28.79} \\
                  & NDCG@10 &0.0586& \textbf{0.0710}\red{21.16} & 0.0531\blue{9.39} & 0.0548\blue{6.48} & 0.0524\blue{10.58} \\
                  & NDCG@20 &0.0655& \textbf{0.0760}\red{16.03} & 0.0618\blue{5.65} & 0.0647\blue{1.22} & 0.0611\blue{6.72} \\ \hline
\multirow{6}{*}{Food} & HR@5 &0.0443& \textbf{0.0742}\red{67.49} & 0.0566\red{27.77} & 0.0472\red{6.55} & 0.0517\red{16.70} \\
                  & HR@10 &0.0597& \textbf{0.0972}\red{62.81} & 0.0747\red{25.13} & 0.0686\red{14.91} & 0.0701\red{17.42} \\
                  & HR@20 &0.0753& 0.1090\red{44.75} & 0.0980\red{30.15} & \textbf{0.1114}\red{47.94} & 0.1043\red{38.51} \\
                  & NDCG@5 &0.0286& \textbf{0.0492}\red{72.03} & 0.0348\red{21.68} & 0.0308\red{7.69} & 0.0321\red{12.24} \\
                  & NDCG@10 &0.0326& \textbf{0.0567}\red{73.93} & 0.0409\red{25.46} & 0.0378\red{15.95} & 0.0388\red{19.02} \\
                  & NDCG@20 &0.0366& \textbf{0.0597}\red{63.11} & 0.0471\red{28.69} & 0.0491\red{34.15} & 0.0462\red{26.23} \\ \hline
\multirow{6}{*}{Home} & HR@5 &0.0133& \textbf{0.0192}\red{44.36} & 0.0175\red{31.58} & 0.0144\red{8.27} & 0.0163\red{22.56} \\
                  & HR@10 &0.0223& \textbf{0.0299}\red{34.08} & 0.0281\red{26.01} & 0.0215\blue{3.59} & 0.0231\red{3.59} \\
                  & HR@20 &0.0270& 0.0337\red{24.81} & 0.0330\red{22.22} & \textbf{0.0341}\red{26.30} & 0.0328\red{21.48} \\
                  & NDCG@5 &0.0097& \textbf{0.0122}\red{25.77} & 0.0087\blue{10.31} & 0.0071\blue{26.80} & 0.0089\blue{8.25} \\
                  & NDCG@10 &0.0109& \textbf{0.0157}\red{44.04} & 0.0118\red{8.26} & 0.0082\blue{24.77} & 0.0103\blue{5.50} \\
                  & NDCG@20 &0.0134& \textbf{0.0167}\red{24.63} & 0.0131\blue{2.24} & 0.0110\blue{17.91} & 0.0126\blue{5.97} \\ \hline
\end{tabular}
\caption{Results of different LLM backbones, the best score is highlighted in \textbf{bold}. The performance difference (\%) with SOTA baselines is highlighted with {\color{RoyalBlue}{blue}} and {\color{Salmon}{orange}}.}
\label{tab:generalization}
\end{table*}

\section{Generalizability of the Method}
To evaluate the generalizability of our method, we conduct experiments using a variety of alternative LLM backbones. Specifically, we test Llama3.2-3B-Instruct \cite{llama3-2}, Gemma2-9B-Instruct \cite{gemma} and Command-R-7B \cite{commandr}, which represent a diverse set of model architectures and parameter scales. These backbones were selected to provide a robust and comprehensive evaluation of our method. 
The results, presented in Table \ref{tab:generalization}, demonstrate that our method remains effective on the Food dataset, but exhibits uneven performance on the other two datasets. In both the Food and Home datasets, the HRs are generally above the baselines, but the NDCGs are noticeably lower, indicating that while relevant items are being recommended, their ranking positions are suboptimal. We speculate that these discrepancies stem from inherent differences in the LLMs themselves, as their knowledge bases and reasoning abilities vary significantly, which can impact their ability to process and prioritize diverse types of information.

\end{document}